\documentclass{aa}
\usepackage{graphicx}
\begin{document}
   \title{VGCF Detection of Galaxy Systems at Intermediate Redshifts}

   \author{R. Barrena\inst{1} 
          \and
           M. Ramella\inst{2}
          \and
           W. Boschin\inst{3,4}
          \and
           M. Nonino\inst{2}
          \and
           A. Biviano\inst{2}
          \and
           E. Mediavilla\inst{1}
         }

   \offprints{R. Barrena, \email{rbarrena@iac.es}}

   \institute{Instituto de Astrof\'{\i}sica de Canarias, V\'{\i}a
   L\'actea s/n, E-38200 La Laguna, Tenerife, Spain\\
         \and
              INAF, Osservatorio Astronomico di Trieste, via G.B. Tiepolo 11, 
	      I-34131 Trieste, Italy\\
         \and
	      INAF, Telescopio Nazionale Galileo,
              Roque de Los Muchachos, PO box 565, 38700 Santa Cruz de La Palma, 
	      Spain\\
         \and
              Dipartimento di Astronomia of the Universit\`a degli Studi di Trieste, via G.B. Tiepolo 11, 
	      I-34131 Trieste, Italy\\
              }

   \date{Received  / Accepted }

\abstract{We continue the development of our $Voronoi$ $Galaxy$ 
$Cluster$ $Finder$ (VGCF) technique by applying it to galaxy catalogs 
obtained with $B$ and $R$ band observations of four high galactic latitude 
fields of $0.5 \times 0.5$ square degrees each. These fields are deep 
($R_{\bf{lim}} \sim 23$, $B_{\bf{lim}} \sim 26$) and partially overlap 
the $Palomar$ $Distant$ $Cluster$ $Survey$ (PDCS) fields at 0$^{\bf{h}}$ 
and 2$^{\bf{h}}$. We run the VGCF also on the original $V$ and $I$ bands 
PDCS galaxy catalogs.\\
We identify a total of 48 clusters that are particularly reliable
being detected in at least two bands. The analysis of color-magnitude
diagrams and, in a few cases, spectroscopic observations allow us to
further increase the reliability of 25 of the 48 clusters. For these
26 clusters we also estimate redshifts that fall in the approximate
range $0.2 < z < 0.6$.\\
We detect 41 VGCF clusters within the strict limits of the PDCS fields
at 0$^{\bf{h}}$ and 2$^{\bf{h}}$. The PDCS catalog for the same
regions consists of 28 clusters. The two catalogs have 20 clusters in
common.  These clusters together with the remaining PDCS and VGCF
clusters lead to a total number of 46 ``independent'' clusters.  The
total number of clusters is therefore 20\% larger than the number of
VGCF clusters and more than 60\% larger than the number of PDCS
clusters.  These results confirm a) that the VGCF is a competitive
algorithm for the identification of optical clusters, and b) that a
combined catalog of matched-filter and VGCF clusters constitutes a
significant progress toward a more complete selection of clusters from
bidimensional optical data.
}  \maketitle

\section{Introduction}

Clusters of galaxies play several roles in modern astrophysics.  For
example, they are laboratories where to test theories of galaxy
formation and evolution.  They are probes of the large scale
structure.  They are gravitational telescopes enabling us to see
distant objects that would be otherwise not detectable. Clusters also
provide powerful constraints on cosmological models.  It is therefore
not surprising that so many efforts have been, and are being spent to
identify clusters of galaxies. Ideally one would like to identify
clusters with a well-defined selection criterion over the widest
redshift range including clusters with the widest range of properties.


In practice, there are two main methods to detect clusters. The first,
classical technique, based on optical/NIR observations, consists in
detecting clusters by discovering galaxy concentrations with respect
to the mean galaxy background/foreground density. The second one,
based on observations on the soft X--ray band (0.1 - 10 keV), consists
in detecting clusters by observing the X--ray emission from the hot
intracluster medium.

Both optical and X--ray selected samples have advantages and
difficulties. In the last 10--15 years X--ray cluster surveys had a
strong development (see, e.g., Rosati et al. \cite{Rosati98}, Boschin
\cite{Boschin02}, Moretti et al. \cite{Moretti04} for recent surveys)
mainly thanks to ROSAT, XMM and Chandra satellites. The main advantage
of X--ray surveys is that clusters look sharper in the X--ray sky than
in the optical sky. In fact, the background is low in X--ray
observations and the X--ray cluster emission is proportional to the
square of the local gas density.

Unfortunately low mass systems, whose X--ray emission is generally
weak, are very difficult to detect, even at low
redshifts. Furthermore, there may be clusters of similar optical
richness but different ICM histories leading to different X--ray
luminosities (Gilbank et al. \cite{Gilbank04}; Donahue et
al. \cite{Donahue01}). In view of these problems it is interesting to
identify clusters from the galaxy distribution in optical images.

In this paper we are concerned with optical identifications.  They
have a big advantage with respect to X--ray identifications: the
availability of a large number of big aperture ground telescopes with
high quantum efficiency detectors.

In fact, in the last decade several deep imaging cluster surveys have
been performed. Notable examples include the Palomar Distant Cluster
Survey (PDCS, Postman et al. \cite{{PLG96}}), the ESO Imaging
Survey (EIS, Nonino et al. \cite{Nonino99}; Olsen et
al. \cite{Olsen99}), the drift-scan survey of Zaritsky et
al. \cite{Zaritsky97}, the KPNO Deeprange (Postman et
al. \cite{Postman02}). These surveys are typically designed to find
rich clusters at high redshift within deep galaxy surveys of
relatively small sky coverage.

Wide field galaxy surveys based on the digitization of photographic
plates have produced (e.g. Dalton et al. \cite{Dalton97}; Lumsden et
al.  \cite{Lumsden92}), and are still producing (e.g. Lopes et al.
\cite{Lopes04}), impressive results.  However, modern wide field
imaging is becoming increasingly common since large format CCD cameras
are now available at several telescopes. These new instruments allow a
systematic search of medium-high redshift galaxy clusters within
photometric catalogs covering wide areas of the sky.

The Sloan Digital Sky Survey, SDSS (\verb+http://www.sdss.org/+), has
been used so far to produce two cluster catalogs (Goto et
al. \cite{Goto02}; Bahcall et al.  \cite{Bahcall03}).  Other notable
cluster catalogs based on wide-field surveys are the Stanford Cluster
Search (Willick et al. \cite{Willick01}, StaCS), the Toronto
Red-Sequence Cluster Survey (Gladders \& Yee \cite{Gladders00};
Gladders \& Yee \cite{Gladders05}), and the Las Campanas Distant
Cluster Survey (Gonzalez et al. \cite{Gonzalez01}, LCDCS).

Since there is a significant background in optical when searching for
clusters (especially those at intermediate and high redshifts), galaxy
systems are identified within galaxy surveys using several techniques
which selectively suppress the background. They are mainly based on
two different algorithms (see Kim et al. \cite{Kim02}): the
matched-filter algorithm (Postman et al. \cite{PLG96}), hereafter
PLG96) and the Voronoi tessellation algorithm (Ramella et al. 
\cite{Ramella01}).
 
Among other techniques there are the cut-and-enhance method (Goto et
al. \cite{Goto02}) and the cluster identification based on the
detection in color space of the red sequence of early-type galaxies at
the core of (rich) clusters (Gladders \& Yee \cite{Gladders00}).

In a previous paper (Ramella et al. \cite{Ramella01}, hereafter R01)
we propose the VGCF, a procedure based on the Voronoi tessellation to
identify clusters within photometric galaxy catalogs. As we specify in
R01, the procedure is meant to complement other techniques, in
particular the matched filter of PLG96.  The matched filter
algorithm is powerful because it is an optimal filtering technique
that can detect clusters that are weak overdensities of the angular
galaxy density distribution. No technique is perfect: the matched
filter algorithm can miss clusters because of its model-dependency and
because of the smoothing of the data that "washes out" small
structures in the periphery of large overdensities.  The VGCF has
among its strengths the fact of being model independent and the lack
of smoothing. On the other hand, some of the matched filter clusters
would never be detectable by the VGCF because of their low
signal-to-noise ratio (SNR) and/or because of their lack of a ``cusp''
in their density profile. The combined use of both techniques should
lead to a significant improvement on the completeness of optically
selected cluster catalogs.

Here we follow up on our previous work on the VGCF (R01) and discuss
its performances on our own multi-band photometric data, as well as on
PDCS data.

The plan of this paper is as follows. In \S 2 we describe observations
and image reduction techniques. In \S 3 we introduce the cluster
detection technique and present our catalog of clusters in \S 4. In
\S5 we study color properties of clusters and, for a subsample, we
also estimate photometric redshifts. In \S 6 we compare our VGCF
catalog to the PDCS catalog. We present the summary of our work in \S
7.

Throughout this paper $\Omega_m=0.3$, $\Omega_\Lambda=0.7$ and $H_0 =
70$ km s$^{-1}$ Mpc are used.

\section{Observations and Galaxy Catalogs}
\label{sec:observations}

Our aim is to test the performances of the VGCF (R01) on new
multi-band observations where images could help our understanding of
the nature of the VGCF detections. At the same time we want to
improve the comparison with the PDCS catalog, generated with a
matched-filter algorithm, in order to verify that we can significantly 
increase the completeness of cluster catalogs by using both search methods.

To this purpose we select 4 pointings for the Wide Field Camera (WFC)
mounted at the prime focus of the 2.5m Isaac Newton Telescope (INT),
located at the Roque de Los Muchachos Observatory, La Palma.  We
carried out observations from October 12 to October 14, 1999 using the
$B_{\bf{H}}$ and $R_{\bf{H}}$ Harris filters. The log of our
observations is in Table \ref{tab:obslog}. We list the coordinates of
our pointings in Table \ref{tab:points}.  The field of view of the WFC
is $34 \times 34$ square minute and the scale is 0.333$''$/pixel.
With this field of view, our pointings allow us to observe 12 PDCS
clusters (PLG96) located in the PDCS fields at 0$^{\bf{h}}$ and
2$^{\bf{h}}$ (hereafter PDCS0 and PDCS2, respectively). The properties
of these clusters are representative of those of the whole PDCS
catalog.  The partial overlap of two of our pointings allows us to
check the consistency of our photometric and astrometric
reductions. It also allows us to check of the stability of our
clustering analysis.

We perform multiple integrations in the direction of each pointing. We
offset each integration by about 20$''$. The dithering pattern of the
integrations allows us to create a ``supersky'' that we use to correct
our images from fringing patterns (e.g. Gullixson
\cite{Gullixson92}). The dithering also allows us to clean cosmic rays
and to avoid gaps between the CCDs of the WFC in the final images.

We perform a standard reduction of our WFC observations with the
IRAF\footnote{IRAF: Image Reduction and Analysis Facility distributed
by National Optical Astronomy Observatories.} package. We reduce
separately each CCD frame. We also find astrometric solutions for each
frame separately using the USNO\footnote{United States Naval
Observatory, version 1.0} catalogs.

We coadd the different dithering frames into a single image using our
own routines. Based on the USNO catalogs, the accuracy of our
astrometry is $\sim 0.5''$ over the full field of the WFC without
significant aberrations in the PSF. The mean ellipticity of the PSF is
0.11 over the whole WFC field. Residual PSF elongations have no
preferred direction.

\begin{table}
\caption{Log of the observations with WFC@INT.}
\label{tab:obslog}
\centering
\begin{tabular}{llccc} 
\hline \hline
 Date & Field & $t_{\bf{exp}}$ (s) & Seeing & Airmass \\
\hline
\multicolumn{5}{c}{$B_{\bf{H}}$ band} \\
 99-10-13 & F0028+0515 & 10000 & $1.6''$ & 1.8-1.1 \\
 99-10-14 & F0027+0555 & 10000 & $1.6''$ & 1.2-1.1 \\
 99-10-13 & F0228+0115 & 10000 & $1.7''$ & 1.4-1.2 \\
 99-10-14 & F0226+0106 & 10000 & $1.1''$ & 1.6-1.1 \\
\hline
\multicolumn{5}{c}{$R_{\bf{H}}$ band} \\
 99-10-12 & F0028+0515 & 3600 & $1.4''$ & 1.8-1.2 \\
 99-10-14 & F0027+0555 & 4200 & $1.6''$ & 1.9-1.2 \\
 99-10-12 & F0228+0115 & 4800 & $1.2''$ & 1.7-1.3 \\
 99-10-13 & F0226+0106 & 6000 & $1.4''$ & 1.5-1.1 \\
\hline
\end{tabular}
\end{table}

\begin{table}
\caption{Coordinates and areas of INT pointings and PDCS fields.}
\label{tab:points}
\centering
\begin{tabular}{lcccc} 
\hline \hline
 Field & $\alpha_{\bf{2000}}$ & $\delta_{\bf{2000}}$ & \multicolumn{2}{c}{Area (sq. deg.)} \\
\hline
  &  &  & B$_{\bf{H}}$ & R$_{\bf{H}}$ \\
 F0028+0515 & 00:28:50 & 05:15:20 & 0.277 & 0.284 \\
 F0027+0555 & 00:27:25 & 05:55:35 & 0.296 & 0.321 \\
 F0228+0115 & 02:28:05 & 01:15:30 & 0.285 & 0.303 \\
 F0226+0106 & 02:26:25 & 01:06:20 & 0.325 & 0.294 \\
\hline 
            &          &          & V$_4$ & I$_4$ \\
 PDCS $0^{\bf{h}}$ & 00:29:11 & 05:31:55 & 1.074 & 0.970 \\
 PDCS $2^{\bf{h}}$ & 02:28:33 & 00:55:45 & 1.120 & 0.927 \\
\hline
\end{tabular}
\end{table} 

Because our three nights at the INT were not photometric, we calibrate
our WFC images with observations obtained with the 1m Jacobus Kaptein
Telescope (JKT; Roque de los Muchachos Observatory) during the night
of September 22, 2000. We select one pointing for each of our INT
frames. The criterion for the selection of the pointings is to have
enough non saturated stars and galaxies in the WFC images in the
magnitude range $R=17-19$. Exposure times of 30 min at the JKT give
good SNR images in this magnitude range.  We typically observe 20 --
30 objects (both stars and galaxies) per field.

We perform \verb+AUTOMAG+ aperture photometry on standard stars using
the $SExtractor$ photometry package (Bertin \& Arnouts
\cite{Bertin96}). With this process we obtain the instrumental
magnitudes ($m_{\lambda,\bf{ins}}$).  The long exposure time of the
scientific images (of the order of 10000 s) does not allow an
estimate of an airmass for the coadded images. For this reason, we can
only determine a reduced transformation for photometric calibrations
by neglecting the color term. The reduced transformations have the
form $m_\lambda = m_{\lambda , \bf{ins}} + b_\lambda$; where $m_\lambda$ is
the tabulated magnitude of the star in each band $\lambda$ and
$b_\lambda$ is the offset magnitude.

We finally identify galaxies in our $B_{\bf{H}}$ and $R_{\bf{H}}$
images and measure their magnitudes with the {\it SExtractor} package.
To these catalogs of galaxies, we add those in the $V_{\bf{4}}$ and
$I_{\bf{4}}$ bands kindly provided to us by M. Postman.

As a final step, we transform all magnitudes into the Johnson-Cousins
system (Johnson \& Morgan \cite{Johnson53}; Cousins \cite{Cousins76}).

For $B$ and $R$ we use $B = B_{\bf{H}} + 0.13$ and $ R = R_{\bf{H}} $,
derived from the Harris filter characterization
(\verb+http://www.ast.cam.ac.uk/~wfcsur/technical/+
\newline \verb+/photom/colors/+) and assuming a $B-V \sim 1.0$ for E-type 
galaxies (Poggianti \cite{Poggianti97}).

For $V$ and $I$ we use the transformations given in PLG96.

Because the PDCS fields are at high galactic latitude, galactic
extinction is low. We estimate $A_{\bf{B}} \sim 0.09$, $A_{\bf{V}}
\sim 0.06$, $A_{\bf{R}} \sim 0.04$ and $A_{\bf{I}} \sim 0.03$ from
Burstein \& Heiles (\cite{Burstein82}) reddening maps.

In conclusion, we have four galaxy catalogs, one for each
(Johnson-Cousins) photometric band $B, V, R$, and $I$. We list in
Table \ref{tab:compmag} completeness magnitude, $m_{\bf{c}}$ ,
and limiting magnitude $m_{\bf{lim}}$, of these 
four catalogs. Completeness and limiting magnitudes correspond
to 5-$\sigma$ and 3-$\sigma$ source detections respectively.

As a part of this project we also obtain spectroscopic data using the
Device Optimized for the Low Resolution (DOLoRes) at the Telescopio
Nazionale Galileo (TNG). We use DOLoRes both in Multi-Object
Spectroscopy (MOS) mode and in Long Slit mode. The size of the MOS
fields is $6 \times 9$ arcmin$^2$ on a Loral 2048$\times$2048 CCD with
15$\mu$ pixels. We use the LR-B grism providing a resolution of
2.8$\,$\AA/pix and a wavelength coverage from 3000$\,$\AA \ to
8800$\,$\AA. With integration times of 90 min and with an average
seeing of 1.5$''$, we obtain spectra for 142 targets with 4 masks and
slits $1.1''$ wide. SNR are in the range 5 to 10 for targets with
magnitudes as deep as $R \simeq 21$. We use the same set-up also in
Long Slit mode, although in this case we widen the slit up to 1.5$''$
because of the worse seeing ($\sim 1.5''$). Total exposure times in
Long Slit mode are of 45 min.

We reduce the TNG and CFHT data using the standard procedures of the
IRAF package. We perform cosmic ray rejection, sky subtraction,
aperture extraction and wavelength calibration. In most cases the SNR
ranges from 5 to 10, or even exceeds these values.  We measure the
redshift using the cross-correlation technique $xcsao$ (Tonry \& Davis
\cite{Tonry79}) implemented in the RVSAO package (developed at the
Smithsonian Astrophysical Observatory Telescope Data Center). We use
as templates the spectra of Elliptical, S0, Sa, Sb, Sc and Irregular
type galaxies taken from Kinney et al. (\cite{Kinney96}).

\begin{table}
\caption{Completeness and limiting magnitudes}
\label{tab:compmag}
\centering
\begin{tabular}{lcccc} 
\hline \hline
 Field  & \multicolumn{2}{c}{$m_{\bf{c}}$} & \multicolumn{2}{c}{$m_{\bf{lim}}$} \\
\hline
 & B & R & B & R \\
 F0028+0515 & 24.4 & 21.8 & 26.0 & 23.1 \\
 F0027+0555 & 24.5 & 22.9 & 26.2 & 24.2 \\
 F0228+0115 & 24.4 & 22.3 & 25.8 & 23.8 \\
 F0226+0106 & 23.4 & 21.3 & 24.7 & 23.3 \\
\hline 
            & V    & I    & V    & I \\
 PDCS $0^{\bf{h}}$ & 23.5 & 21.0 & 24.8 & 22.8 \\
 PDCS $2^{\bf{h}}$ & 23.5 & 21.2 & 25.1 & 23.3 \\
\hline
\end{tabular}
\end{table} 

\section{Detecting clusters with the VGCF}
\label{sec:vgcf}

In order to identify clusters of galaxies in our galaxy catalogs we
use the VGCF. This technique is described in detail in R01. Here we
briefly summarize the main characteristics of the VGCF.

The VGCF uses the Voronoi tessellation in order to assign to each
object a local density given by the inverse of the area of the Voronoi
tessel of that point. We remind that a Voronoi tessellation of a
two-dimensional distribution of galaxies is a unique plane partition
into convex cells, each of them containing one, and only one object.

Then the VGCF determines the background density of objects by fitting
the low-density end of the observed integral density distribution.

The VGCF compares the observed density distribution with the empirical
density distribution expected for a Poissonian distribution of points
having the same density as derived from the fit to the low-density end
of the observed integral density distribution.

According to Kiang (\cite{Kiang66}), the expected background
distribution has the form:

\begin{displaymath} dp(\tilde{a}) = \frac{4^4}{\Gamma(4)} \tilde{a}^3
e^{-4\tilde{a}} d \tilde{a} \end{displaymath} 

where $\tilde{a} \equiv a/<a>$ is the cell area in units of the
average cell area $<a>$.  As each cell contains exactly one galaxy,
the corresponding density is the inverse of the cell area $f \equiv
1/a$.

The user establishes a density threshold above the Poissonian
distribution, i.e. a minimum confidence level for significant
overdensities.  Here, as in R01, we set this threshold at the 80\%
level.

The algorithm then defines overdense regions as those composed by
adjacent Voronoi cells with a density higher than the chosen
threshold.  By computing the probability that an overdensity
corresponds to a background fluctuation (see R01), the VGCF discards
overdensities with probabilities greater than a given threshold. We
set this threshold at the 95\% level.  

The VGCF regularizes the shape of the overdense regions. First, it
assumes that all the points inside the convex hull defined by the set
of points belong to the overdensity itself. Next, if fits a circle to
the overdense region and expands it until the mean density inside the
circle is lower than the density of the original region.

The output of the VGCF is a catalog of overdensities, or clusters,
listed with their main characteristics.

We note that the VGCF does not assume density or luminosity profiles
for clusters and does not smooth the data. The VGCF can identify
clusters irrespective of their shape and is only weakly affected by
edge effects and by "holes" in the galaxy distribution (e.g. those
caused by extremely bright stars). The algorithm is fast, and
automatically assigns members to the structures.

Because of the depth of our galaxy catalogs, and the corresponding
high galaxy projected number densities, it is necessary to run the
VGCF in magnitude bins, as in R01.

The choice of the bin width can not be equally optimal for the
detection of all clusters. Following R01, we use bins that are 2
magnitudes wide. In fact, R01 gage their bins for cluster catalogs
covering a redshift range very close to that of our present sample.

We select galaxies within a magnitude bin and run the VGCF over this
sub-sample of galaxies. We then shift faintward the magnitude bin by
0.1 magnitudes and run the VGCF again. The width of the step is the
same as in R01.

We note that we select the first bin starting from fainter magnitudes
than those of the brightest galaxies. This is necessary in order to
have enough counts in the first bin for a correct estimate of the
background galaxy density. The omission of the (few) brightest
galaxies has no impact on our analysis. The brightest and faintest
magnitudes of our binning procedure are summarized in Table
\ref{tab:catbin}. In the end we obtain at least 25 catalogs in each
band.

\begin{table}
\caption{\small Details of the magnitude bins of the VGCF runs.}
\centering
\begin{tabular}{lcccc} 
\hline \hline
 Field$^{\mathrm{a}}$ & First bin & Last bin & First bin & Last bin \\
\hline
  & \multicolumn{2}{c}{Catalog in B band} & 
  \multicolumn{2}{c}{Catalogs in R band}\\
 F0028 & 20.0-22.0 & 22.5-24.5 & 18.0-20.0 & 20.5-22.5 \\
 F0027 & 20.0-22.0 & 22.5-24.5 & 19.0-21.0 & 21.5-23.5 \\
 F0228 & 20.0-22.0 & 22.5-24.5 & 18.0-20.0 & 20.5-22.5 \\
 F0226 & 20.0-22.0 & 22.5-24.5 & 17.5-19.5 & 20.0-22.0 \\
\hline
 & \multicolumn{2}{c}{Cataloger in V band} & 
 \multicolumn{2}{c}{Cataloger in I band} \\
PDCS0 & 18.0-20.0 & 21.5-23.5 & 17.0-19.0 & 19.5-21.5 \\
PDCS2 & 18.5-20.5 & 21.5-23.5 & 17.0-19.0 & 19.5-21.5 \\
\hline
\end{tabular}
\begin{list}{}{}
\item[($^{\mathrm{a}}$)] F0028, F0027, F0228 and F0226 correspond to F0028+0515, F0027+0555, 
F0228+0115 and F0226+0106 fields, respectively
\end{list}
\label{tab:catbin}
\end{table} 
  
In order to ``merge'' into a single cluster sequences of overlapping
overdensities identified in different magnitude bins (in the same
band), we use similar criteria to those indicated by R01. The slight
variations are dictated by the different data sets at our disposal.
In practice, we proceed as follows.

We consider a set of overdensities a cluster if there are at least 
4 overdensities that: a) have centers that are closer than 15$''$ from
the corresponding overdensity in the next magnitude bin, b) are
replicated within at least 4 magnitude bins that are adjacent or at
most with a total of one bin missing. We then require the cluster to
be identified in at least two photometric bands with centers separated
by less than 15$''$.

By applying these criteria to the VGCF runs in magnitude bins on our
$BVRI$ catalogs we obtain a final list of 48 clusters.

\section{The cluster catalog}
\label{sec:clus_catalog}

\begin{table*}
\caption{Cluster Catalog.}
\label{tab:clust0y2}
\centering
\begin{tabular}{ccccccccccl}
\hline \hline
ID & $C_{\bf{B}}$ & $C_{\bf{V}}$ & $C_{\bf{R}}$ & $C_{\bf{I}}$ & $\alpha_{\rm 2000}$ &
$\delta_{\rm 2000}$ & Radius & $\overline{z}_{\bf{phot}}$ & $z_{\bf{PDCS}}$ & Remarks $^{\mathrm{a}}$ \\
 & & & & & (hh:mm:ss) & ($^{\circ}$:$'$:$''$) & ($''$) & & & \\
\hline
\multicolumn{11}{c}{Field F0028+0515} \\
V01 & 0    & 6.79 &  5.35 & 0	 & 00:28:31.7 & 05:18:18 &  97 &  --   & 0.6 & A, P08 \\
V02 & 5.42 & 5.13 &  0    & 0	 & 00:28:37.0 & 05:07:48 & 119 &  --   & 0.6 & A, P03 \\
V03 & 0    & 5.73 &  0    & 6.41 & 00:29:02.9 & 05:01:30 & 113 &  0.20 & 0.6 & A, P01 \\
V04 & 5.92 & 7.78 & 10.08 & 8.43 & 00:29:13.0 & 05:08:13 & 155 &  --   & 0.5 & A, P04 ($z=0.550$) \\
V05 & 5.67 & 5.29 &  0    & 6.06 & 00:29:13.7 & 05:26:56 & 108 &  0.75 & --  & A \\
V06 & 0    & 5.00 &  4.47 & 5.66 & 00:29:16.3 & 05:18:14 &  94 &  0.43 & --  & A, Confirmed ($z=0.366$) \\
V07 & 5.72 & 5.16 &  0    & 5.68 & 00:29:31.2 & 05:03:40 & 115 &  0.30 & 0.4 & A, P02 \\
V08 & 6.96 & 5.50 &  4.95 & 7.42 & 00:29:51.6 & 05:12:29 & 158 &  --   & 0.4 & A, P06 \\
\hline
\multicolumn{11}{c}{Field F0027+0555} \\
V09 & 8.00  & --    &  7.12 & --   & 00:26:29.5 & 05:41:17 &  97  & --   & --   & B \\
V10 & 11.68 & --    & 10.58 & --   & 00:26:41.3 & 05:47:38 &  90  & 0.61 & --   & B \\
V11 & 7.00  & 0     &  5.67 & 0    & 00:27:30.5 & 05:56:46 &  68  & --   & --   & A \\
V12 & 0     &  5.66 &  5.58 & 0    & 00:27:55.0 & 05:59:20 & 104  & --   &--    & A, inside V13 \\
V13 & 11.26 & 10.03 &  9.77 & 9.81 & 00:27:53.5 & 05:57:11 & 284  & 0.32 & 0.35 & A, P12 \\
\hline
\multicolumn{11}{c}{Field F0228+0115} \\
V14 & 5.82 & 0    &  6.35 & 0    & 02:26:56.4 & 01:03:14 & 101 &  --   & --   & A, inside V33 \\
V15 & 0    & 5.06 &  6.00 & 0    & 02:27:08.9 & 01:03:25 & 122 &  --   & --   & A \\
V16 & 7.42 & 7.00 &  0    & 4.62 & 02:27:20.9 & 01:09:18 & 119 &  0.31 & --   & A \\
V17 & --   & 8.50 &  9.35 & 7.22 & 02:27:42.2 & 00:59:24 & 148 &  0.32 & --   & A \\
V18 & 5.66 & 8.94 &  8.92 & 8.14 & 02:27:48.0 & 01:13:44 & 194 &  --   & --   & A \\
V19 & 5.00 & 6.94 &  6.93 & 7.60 & 02:27:53.0 & 01:04:52 & 191 &  0.19 & 0.2  & A, P23 \\
V20 & 6.00 & 4.47 &  5.20 & 0    & 02:28:15.6 & 01:13:55 &  83 &  0.47 & --   & A \\
V21 & 8.05 & --   & 11.01 & --   & 02:28:23.5 & 01:30:40 & 202 &  0.44 & --   & B \\
V22 & 8.00 & --   &  8.08 & --   & 02:28:27.1 & 01:27:54 &  65 &  --   & --   & A, inside V21 \\
V23 & --   & 6.36 &  7.00 & 7.00 & 02:28:32.4 & 00:57:18 & 101 &  0.34 & 0.45 & A, P20 \\
V24 & 5.20 & --   &  5.00 & --   & 02:28:43.4 & 01:28:05 &  97 &  0.28 & --   & A \\
V25 & 9.26 & 6.00 &  6.68 & --   & 02:28:49.7 & 01:23:13 & 288 &  --   & 0.2  & A, P28 \\
\hline
\multicolumn{11}{c}{Field F0226+0026} \\
V26 & 5.82 & --   &  4.95 & --   & 02:25:19.0 & 00:49:37 & 104 &  --   & --   & B \\
V27 & 6.20 & --   &  7.89 & --   & 02:25:43.7 & 00:49:55 & 104 &  --   & --   & B \\
V28 & 6.38 & --   &  6.32 & --   & 02:25:47.3 & 01:07:05 & 108 &  --   & --   & B \\
V29 & 6.10 & --   & 12.49 & --   & 02:26:06.5 & 01:11:02 & 223 &  --   & --   & B \\
V30 & 7.57 & 8.96 & 18.01 & 6.93 & 02:26:25.7 & 01:00:40 & 248 &  0.38 & --   & A \\
V31 & 0    & 0    &  5.31 & 6.93 & 02:26:35.3 & 01:12:58 &  86 &  0.12 & --   & A \\
V32 & 5.37 & 6.38 &  4.91 & 6.06 & 02:26:39.6 & 00:57:50 & 133 &  --   & 0.35 & A, P21 \\ 
V33 & 4.44 & 7.84 &  8.23 & 8.27 & 02:26:53.3 & 01:06:29 & 234 &  --   & 0.5  & A, P22 \\
\hline
\multicolumn{11}{c}{PDCS 0 (no overlap with F0028+0515 or F0027+0555)} \\
V34 & & 6.12 & & 6.72 & 00:27:29.3 & 05:33:00 & 101 &  0.32 &  --   & C \\
V35 & & 8.33 & & 6.83 & 00:28:55.2 & 05:48:04 & 248 &  0.33 &  0.35 & C, P10 \\
V36 & & 8.00 & & 8.22 & 00:29:41.3 & 05:50:35 & 245 &  0.36 &  0.35 & C, P11 \\
V37 & & 7.25 & & 6.76 & 00:29:48.5 & 05:58:08 & 130 &  --   &  --   & C \\
V38 & & 6.40 & & 6.03 & 00:30:47.8 & 05:37:44 & 144 &  0.29 &  0.35 & C, P09 \\
V39 & & 6.00 & & 5.77 & 00:30:53.5 & 05:54:58 &  94 &  0.39 &  --   & C \\
V40 & & 8.49 & & 5.30 & 00:30:55.2 & 05:14:49 & 227 &  0.35 &  0.25 & C, P07 \\
V41 & & 5.00 & & 6.00 & 00:31:16.1 & 05:16:37 & 104 &  0.28 &  --   & C \\
\hline
\multicolumn{11}{c}{PDCS 2 (no overlap with F0228+0115 or F0226+0026 fields)} \\
V42 & & 7.00 & &  6.36 & 02:27:25.9 & 00:25:19 & 137 &  --    & 0.35 & C, P14 \\
V43 & & 6.93 & &  6.36 & 02:27:29.8 & 00:39:07 & 112 &  --    & 0.4  & C, P18 \\
V44 & & 5.55 & &  5.13 & 02:28:01.9 & 00:32:35 & 104 &  --    & --   & C \\
V45 & & 6.87 & &  5.73 & 02:28:02.4 & 00:40:52 & 133 &  0.31  & --   & C \\
V46 & & 9.43 & & 11.35 & 02:28:27.8 & 00:31:26 & 126 &  0.46  & 0.5  & C, P16 \\
V47 & & 5.33 & &  6.00 & 02:30:07.9 & 00:46:16 & 104 &  --    & --   & C \\
V48 & & 8.55 & &  6.42 & 02:30:23.0 & 01:09:29 & 169 &  0.26  & 0.4  & C, P24 \\
\hline      
\end{tabular}
\begin{list}{}{}
\item[($^{\mathrm{a}}$)] In the column with remarks, an "A"
corresponds to a cluster identified in an overlap region between our
fields and PDCS0 or PDCS2, a "B" to a cluster identified in a region
outside the PDCS fields, a "C" to a cluster identified in a PDCS
region not overlapping our fields. PDCS counterparts are labeled with
a "P" followed by the cluster ID in PLG96. We also add a remark  
indicating VGCF clusters that have their center within another VGCF cluster.
\end{list}
\end{table*}

We present the catalog of the 48 VGCF clusters in Table
\ref{tab:clust0y2}. We give the cluster ID in Col. (1). In Cols.
(2), (3), (4) and (5) we list the SNRs in the $B$, $V$, $R$ and $I$
bands respectively. We estimate the SNR as the number of members
within the area of the overdensity divided by the square root of
background counts expected within the same area according to the VGCF
background fit. The SNR values in Cols. (2) -- (5) are averages over
all overdensities associated with the cluster (in each band). A ``0''
indicates that the cluster has not been identified in the
corresponding band and a blank space indicates that we have no data at
that position.

In Cols. (6) and (7) we list J2000 right-ascension and declination,
respectively. In Col. (8) we give the angular radius of the cluster
(in arcsec). The value we list is the average of the radii of the
detections in all bands. We list our photometric redshift estimate in
Col. (9). For the clusters with both $BR$ and $VI$ redshift
estimates, the value we give in the table is the average of the two
estimates.  For clusters that have a PDCS counterpart we give the PDCS
redshift estimate in Col. (10). We add comments in Col. (11).

\begin{figure}
\centering
\includegraphics[width=8.8cm]{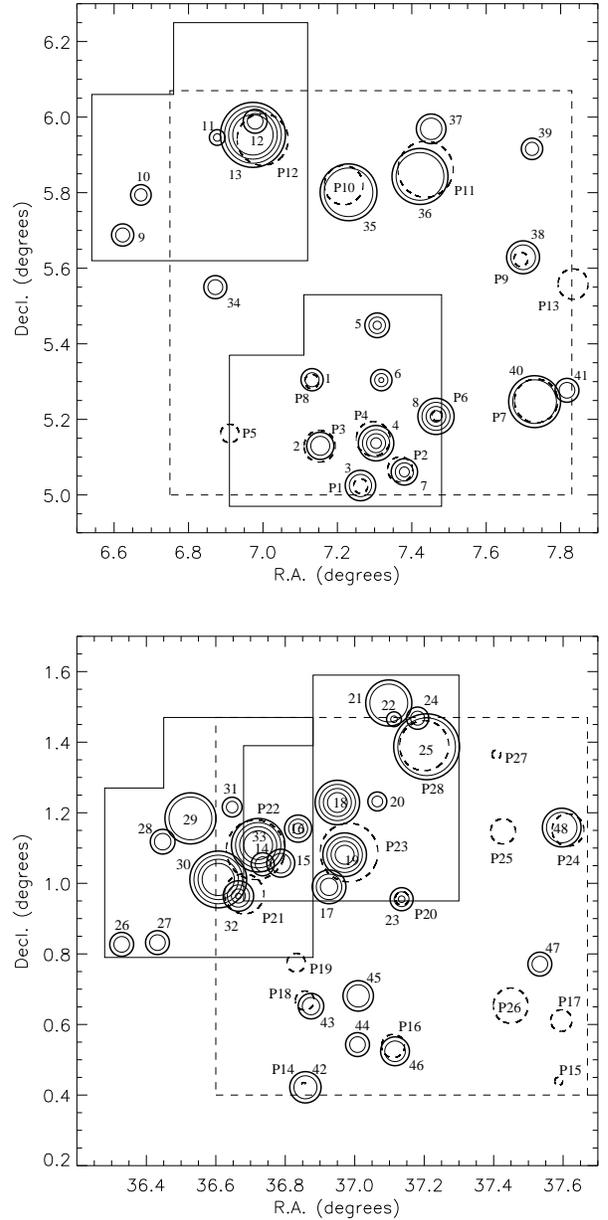}
\caption{VGCF and PDCS clusters. We draw with a thick line our clusters,
with a dashed line PDCS clusters, with a thin line the VGCF detections
in individual bands (the radii of these last detections are arbitrary.
VGCF clusters are labeled with their ID number, PDCS clusters are labeled 
with A  ``P'' followed by the PDCS identification number.}
\label{fig:clust_tot}
\end{figure}

Fig. \ref{fig:clust_tot} is the graphical representation of Table
\ref{tab:clust0y2}.  The dashed square regions represent the PDCS
fields.  The thin solid lines delimit our $B$ and $R$ survey,
i.e. fields \mbox{F0028+0515}, \mbox{F0027+0555}, \mbox{F0228+0115},
and \mbox{F0226+0026}. Circles drawn with thick lines are our
clusters.  Radii correspond to those listed in Table
\ref{tab:clust0y2}.  Circles drawn with thin lines represent
detections in individual bands.  The radius of these circles is
arbitrary. The label next to each circle is the cluster ID number in
Table \ref{tab:clust0y2}. Dashed line circles are the original PDCS
clusters. The radii of these circles are those given by PLG96. We
label these circles with a "P" followed by the PDCS identification
number.

As far as the overlap region between the two fields F0228+0115 and
F0226+0106 is concerned, it provides a check of the reliability of the
detections.  This check is important because of the different
structure of the galaxy distribution in the two fields. In principle,
this difference may cause the VGCF to estimate slightly different
background levels leading to inconsistencies in the overlap region.
Reassuringly, we identify the same set of clusters in both frames
within the overlap region. In Table \ref{tab:clust0y2} the properties
of the clusters in the overlap region are averages of the properties
of the clusters identified within each frame.

\section{Color analysis of cluster candidates}
\label{sec:colorana}

In this section we use our $BVRI$ photometry to find evidences that
our clusters are real physical systems.  We analyze the
Color-Magnitude Diagrams (CMD) of each cluster, and try to identify a
Sequence of Early Type galaxies (ETS).

Virtually all rich clusters have a sequence of early-type galaxies
easily distinguishable in a CMD (Gladders \& Yee \cite{Gladders00},
and references therein). The ETS also gives the possibility of
estimating the redshift of clusters (Gladders \& Yee
\cite{Gladders05})

In order to identify the ETS in our CMD, we start from the ETS of Coma
($z=0.023$) for the ($B-R$,$R$) CMD and from the ETS of A118
($z=0.31$) for the ($V-I$,$I$) CMD. Secker et al. \cite{Secker97} show
that the ETS of Coma follows the relation $(B-R) = (-0.056 \pm 0.002)
R + (2.41 \pm 0.04)$. We fit the ETS of A118 from galaxy members of
this cluster (Busarello et al. \cite{Busarello02}). In this case, the 
red sequence follows the relation $(V-I) = (-0.060 \pm 0.007) I + 
(2.77 \pm 0.09)$.

We then compute the set of ETS that Coma and A118 would have if they
were were moved at the redshifts $z = 0.1, 0.2, 0.3 ... 0.7$.  We
compute the expected ETS by applying to the observed ETS theoretical
evolutionary- and K-corrections for a passively evolving population of
early-type galaxies (Poggianti \cite{Poggianti97}).  We use these
grids of semi-empirical ETS to guide our eye in the detection of
possible sequences in the ($B-R$,$R$) and ($V-I$,$I$) CMD.

Because several of our clusters are rather poor, the impact of
foreground/background interlopers on the identification of ETS may be
significant. For this reason we apply a statistical technique to clean
the CMD. In particular, we perform a statistical subtraction of the
galaxy background using the method proposed by Phelps
(\cite{Phelps97}). He uses his method to clean CMD of star
clusters. We find it to clean quite effectively also the CMD of our
galaxy clusters.

In practice, we proceed as follows. Next to the cluster under study we
select a region (field) of the sky without any cluster identification.
In order to have sufficient statistics, the area of the field is five
times larger than the area containing the cluster under study.

We divide both the CMD of the cluster and of the field into a grid of
20 $\times 20$ square cells with a side of 0.2 magnitude.  We then
eliminate one randomly selected object from each cell of the cluster
CMD every five objects present in the same cell of the CMD of the
field.

When we find an ETS, we use it to derive an approximate indication of
the photometric redshift of the cluster.  The ($B-R$,$R$) ETS
significantly shifts toward redder colors in the redshift range $0.0 <
z < 0.3$.  The ($V-I$,$I$) ETS shifts at the highest rate in the
redshift range $0.3 < z < 0.5$.  Our $B$, $V$, $R$ and $I$ bands are
appropriate for estimating photometric redshifts $z < 0.5$.  Redshift
estimates of clusters with $0.5<z<0.7$ will be quite crude.

\begin{figure*}
\centering
\includegraphics[width=17.5cm]{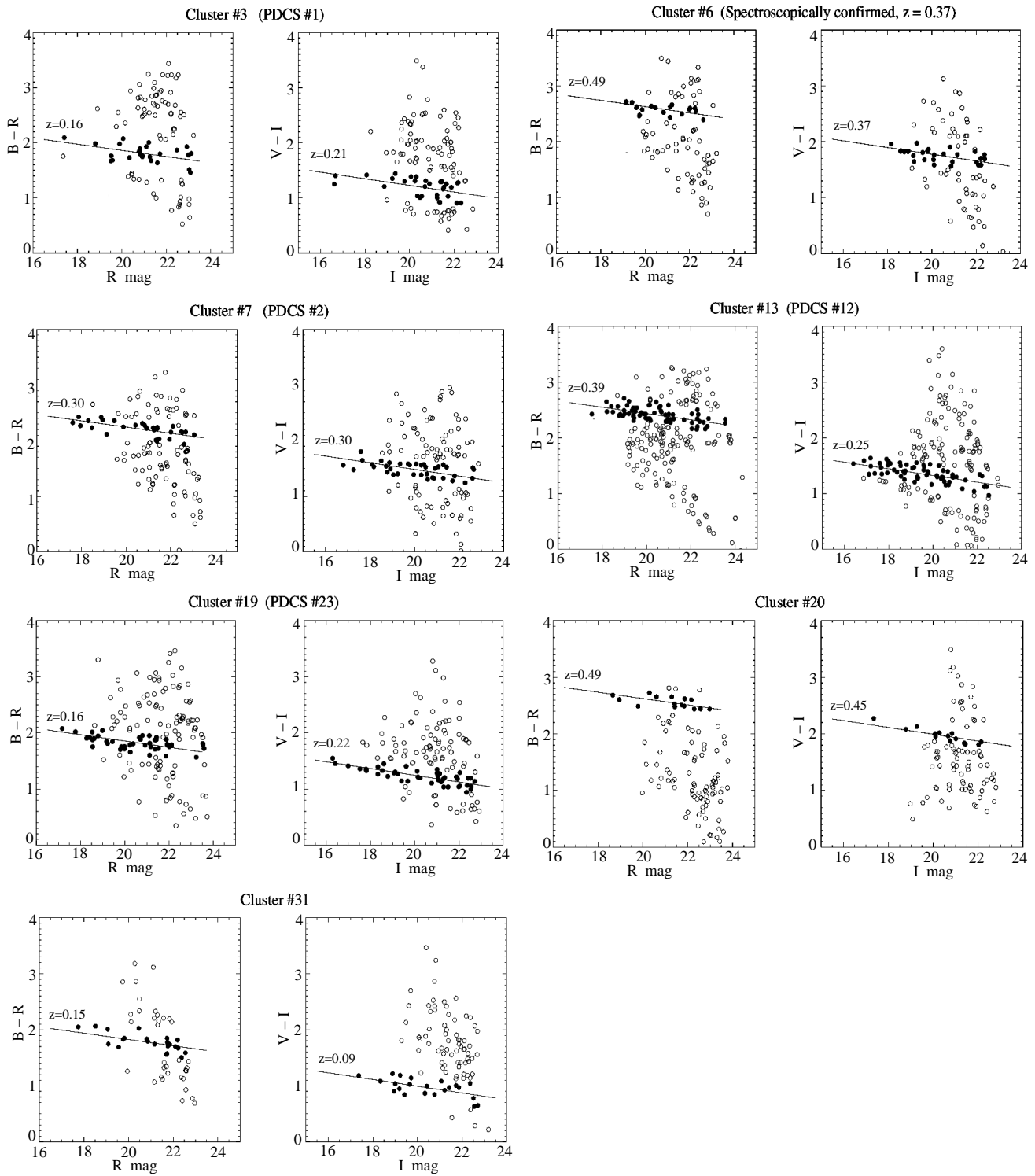}
\caption{($B-R$,$R$) and ($V-I$,$I$) ETS and redshift estimates of
clusters within our survey. This sample includes clusters with ETS in
both ($B-R$,$R$) and ($V-I$,$I$) diagrams.}
\label{fig:clBVRI}
\end{figure*}

\begin{figure*}
\centering
\includegraphics[width=17.5cm, height=23.2cm]{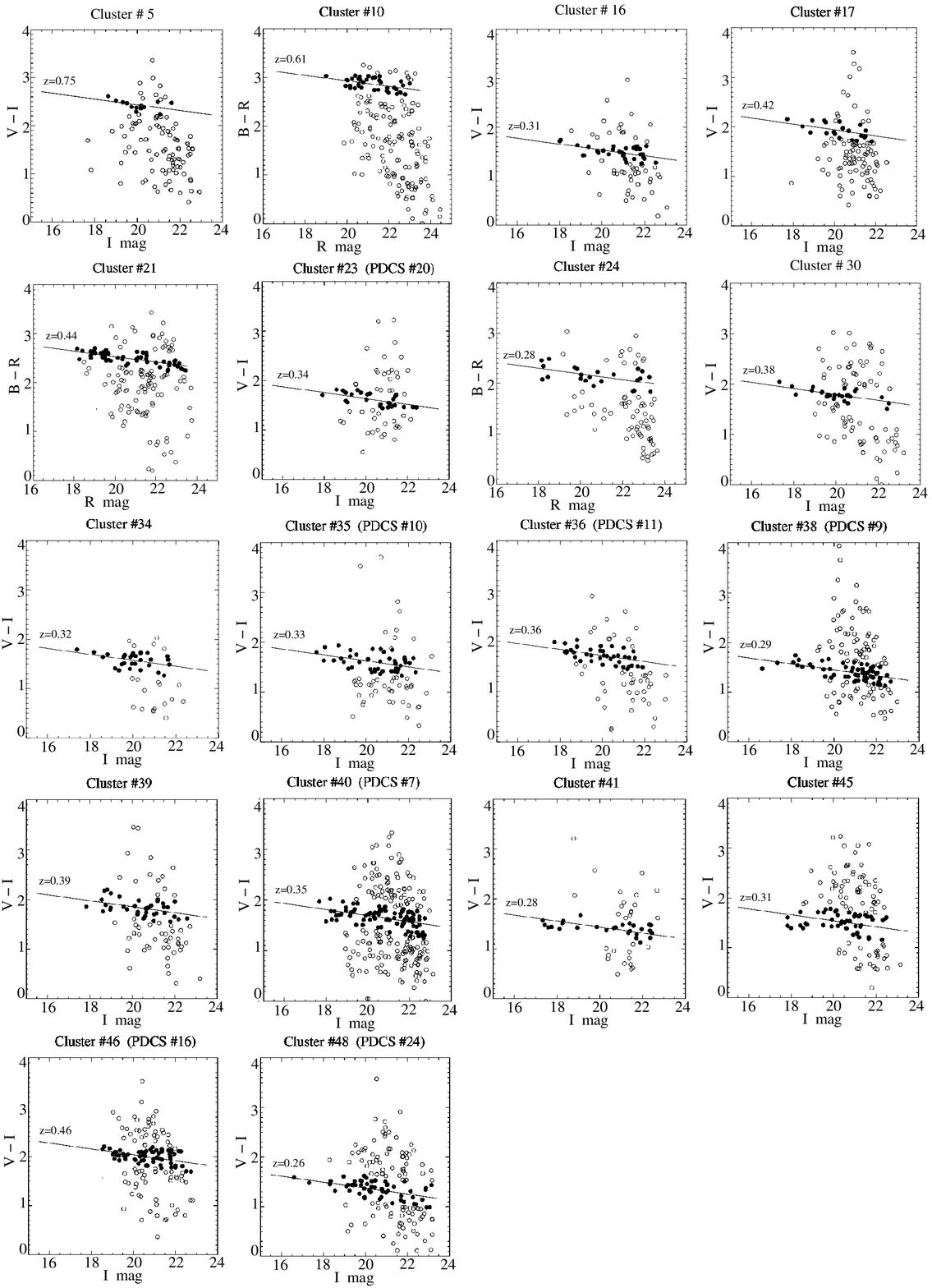}
\caption{ETS and redshift estimates of clusters within our
survey. This sample includes clusters with ETS in only one of either
($B-R$,$R$) or ($V-I$,$I$) diagrams.}
\label{fig:clBRoVI}
\end{figure*}

In order to estimate the photometric redshift we select galaxies
within a color range $\pm 0.3$ mag from the ETS that we identify with
the aid of the semi-empirical grid of ETS.  We then fit a new ETS to
these galaxies. We constrain the slope of the ETS to be the same as
that of the Coma cluster in the ($B-R$,$R$) CMD and of A118 in the
($V-I$,$I$) CMD.

In Fig. \ref{fig:clBVRI} we plot the CMD of the 7 clusters that
present an ETS both in the ($B-R$,$R$) and in the ($V-I$,$I$) CMD. In
Fig. \ref{fig:clBRoVI} we plot the CMD of the 18 galaxy clusters
with evidence of only one ETS in either their ($B-R$,$R$) or
($V-I$,$I$) diagram.

The redshift estimates range from $z_{\bf{ETS}}$ = 0.12 (V31) to $z_{\bf{ETS}}$
= 0.75 (V05). For each cluster in Figs. \ref{fig:clBVRI} and
\ref{fig:clBRoVI} we write the estimated redshift next to the fitted
ETS. We also list all redshifts in Table \ref{tab:clust0y2}. In this
table $z_{\bf{phot}}$ is the average between two redshift estimates if a
cluster has both ($B-R$,$R$) ETS and ($V-I$,$I$) ETS.

\begin{figure*}
\centering
\includegraphics[width=\textwidth]{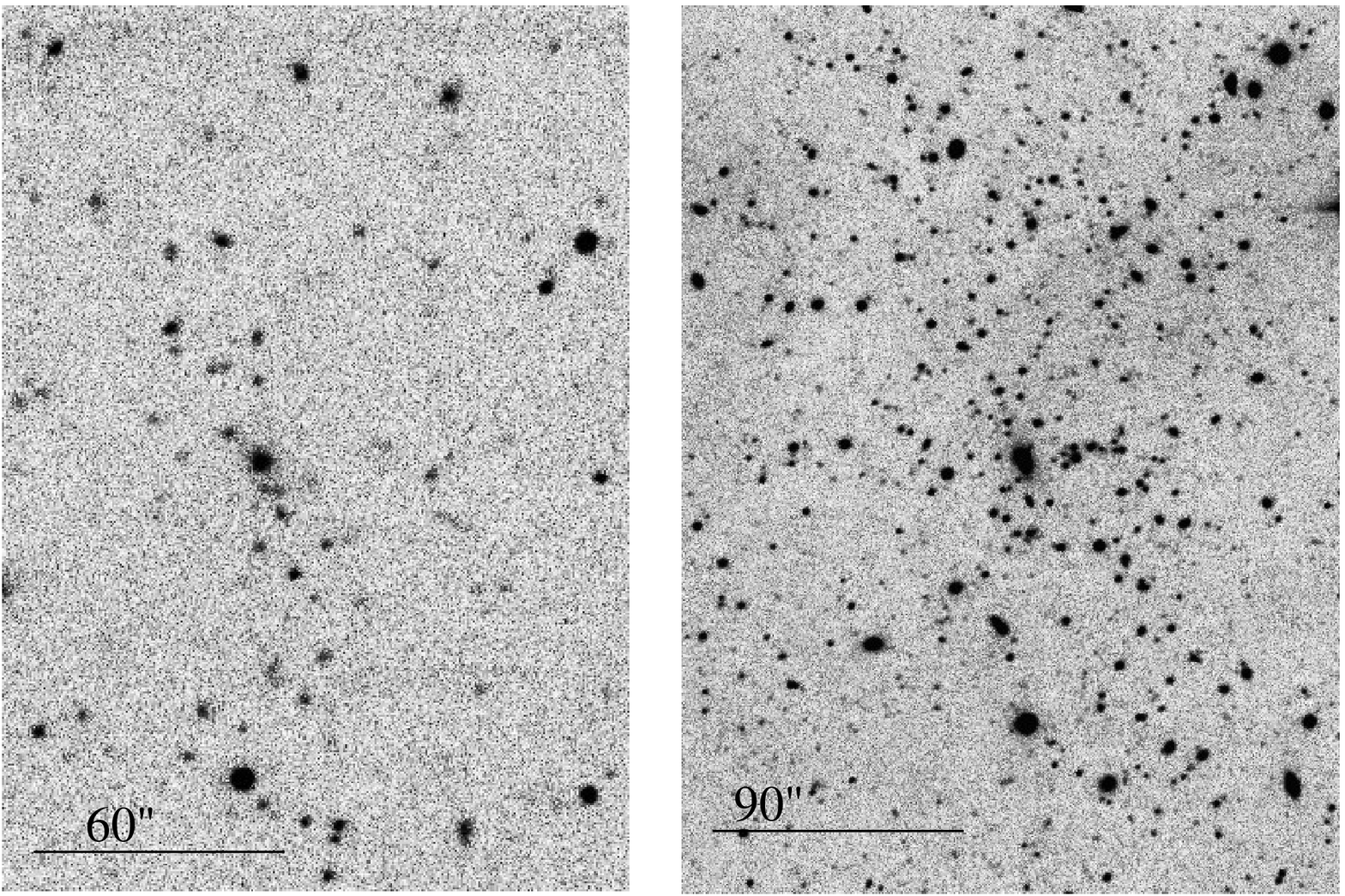}
\caption{$R$ band images of cluster V05 (left panel) and cluster V10
(right panel). ETS reveal clusters at $z = 0.7$ and $z = 0.6$,
respectively.  The central galaxies in these images correspond to
brighter galaxies in the ETS of these systems. North is up and East is
left.}
\label{fig:clust5y10}
\end{figure*}

In Figs. \ref{fig:clust5y10} we show $R$ band images of cluster V05
(left panel) and cluster V10 (right panel).  These two clusters are
our most distant confirmed detections, $z = 0.75$ and $z = 0.6$,
respectively.  The central galaxies in these images correspond to
brightest galaxies in the ETS of these systems.

External and systematic errors are likely to dominate the real
uncertainty of our redshift estimates. Given the accuracy of the
photometry, we can estimate $\delta (B-R)$ and $\delta (V-I)$ to be
about 0.15 magnitude for objects with $R=20$.  These errors produce a
redshift uncertainty of about $\delta z = 0.1$.

An error of $\delta z = $0.1 in the photometric redshifts allows us to
distinguish between "nearby" and "distant" clusters and it is
comparable to the redshift estimate output by the matched-filter
algorithm for PDCS clusters.

\subsection{Spectroscopic confirmation of photometric redshifts}
\label{sec:valred}

We present here follow up observations and archival data we collect in
order to improve/verify our ability to identify real physical
clusters, their possible ETS and the reliability of the photometric
redshift we derive from the ETS in the $BR$ and/or $VI$ data.

We observe clusters V06 and V18 with the Device Optimized for the Low
Resolution (DOLoRes) at the Telescopio Nazionale Galileo (TNG), in
Multi-Object Spectroscopy (MOS) mode.

We retrieve spectra for V04 (P04) from the CFHT archive (see Adami et
al. \cite{Adami00} for details). We reduce the spectra and obtain new
redshift measurements.

Finally, we find within SDSS redshifts of four galaxies, one in each
of the following clusters: V17, V23, V30, and V48.

We list the redshifts we measure in confirmed clusters in Table
\ref{tab:our_redshifts} (cluster V18 is not confirmed
spectroscopically).

We analyze in this section three further PDCS clusters P33 and P36 in
the PDCS field at 9$^{\bf{h}}$ and P63 in the PDCS field at
13$^{\bf{h}}$. For these clusters we have both photometric and
spectroscopic data. We include here the analysis of these clusters
since their data at our disposal are comparable to those of the other
clusters and therefore provide a further check of our procedures of
confirmation and redshift estimation.

$B$ and $R$ photometry of clusters P33, P36, and P63 is from the HiRAC
camera mounted at the 2.5m Nordic Optical Telescope (NOT), located at
Roque de los Muchachos Observatory, La Palma. We carried out these
observations during three nights of February 2000 under very good
photometric and seeing conditions using $B$ and $R$ Johnson filters
(exposure times are 10000 s and 6000 s in the $B$ and $R$ band,
respectively). Spectroscopy is from DOLoRes (used in Long Slit mode)
at the TNG. We measure additional 18 redshifts for P33 from spectra we
retrieve from the CFHT archive.

\begin{figure*}
\centering
\includegraphics[width=\textwidth]{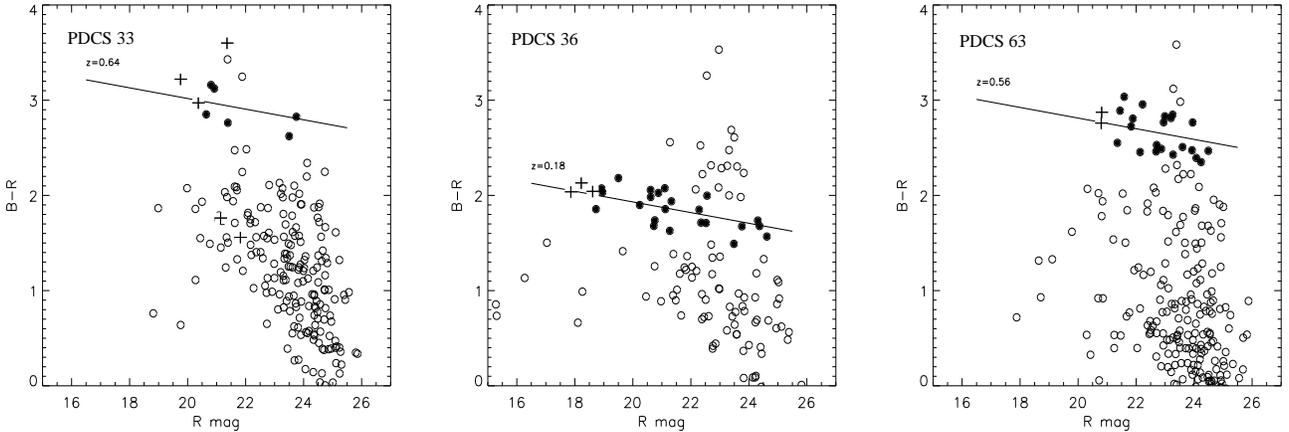}
\caption{($B-R$,$R$) CMD of clusters P33, P36, and P63. The ETS are
linear fits with constant slopes to black filled circles and
crosses. Crosses are cluster members with spectroscopic redshift.}
\label{fig:CMR_PDCS333663}
\end{figure*}

We treat P33, P36, and P63 clusters in the same way as our VGCF
clusters and identify an ETS in each of them. We plot the CDM of these
three clusters in Fig. \ref{fig:CMR_PDCS333663} and list the
redshifts of their members in Table \ref{tab:our_redshifts}.

Finally, in order to test our photometric redshift estimates, we also
use the spectroscopic redshifts of P11, P12, and P23 published in
Holden et al. (\cite{Holden99}).

With these last three redshifts we can compare photometric to
spectroscopic redshifts for 13 clusters.

\begin{table}
\caption{Redshifts of member galaxies.}
\label{tab:our_redshifts}
\centering
\begin{tabular}{lcccc} 
\hline \hline
 ID & R.A.$_{\rm 2000}$ & Dec.$_{\rm 2000}$ & $z$ & $\delta_{\bf{z}}$ \\
    & (hh:mm:ss) & ($^{\circ}$:$'$:$''$) &  & \\
\hline
\multicolumn{5}{c}{Cluster V04 (P04)} \\
  1 $^{(+)}$ & 00:29:13.1 & 05:09:15 & 0.5492 & 0.0004 \\
  2 $^{(+)}$ & 00:29:12.1 & 05:08:32 & 0.5512 & 0.0003 \\
  3 $^{(+)}$ & 00:29:11.0 & 05:09:27 & 0.5509 & 0.0004 \\
  4 $^{(+)}$ & 00:29:10.1 & 05:09:10 & 0.5494 & 0.0003 \\
  5 $^{(+)}$ & 00:29:15.6 & 05:07:11 & 0.5539 & 0.0002 \\
  6 $^{(+)}$ & 00:29:14.4 & 05:07:58 & 0.5504 & 0.0001 \\
  7 $^{(+)}$ & 00:29:11.0 & 05:08:45 & 0.5513 & 0.0001 \\
  8 $^{(+)}$ & 00:29:15.9 & 05:10:03 & 0.5481 & 0.0001 \\
  9 $^{(+)}$ & 00:29:17.7 & 05:10:00 & 0.5498 & 0.0002 \\
\hline 
\multicolumn{5}{c}{Cluster V06} \\
  1 & 00:29:13.7 & 05:17:04 & 0.3678 & 0.0007 \\
  2 & 00:29:16.0 & 05:17:45 & 0.3652 & 0.0002 \\
  3 & 00:29:16.1 & 05:17:56 & 0.3640 & 0.0003 \\
  4 & 00:29:18.6 & 05:19:01 & 0.3671 & 0.0004 \\
  5 & 00:29:09.6 & 05:13:50 & 0.3691 & 0.0002 \\
  6 & 00:29:15.9 & 05:17:49 & 0.3682 & 0.0006 \\
  7 & 00:29:25.6 & 05:17:06 & 0.3619 & 0.0002 \\
  8 & 00:29:15.8 & 05:17:37 & 0.3658 & 0.0001 \\
  9 & 00:29:23.7 & 05:20:33 & 0.3620 & 0.0001 \\
\hline 
\multicolumn{5}{c}{Cluster P33} \\
  1 $^{(*)}$ & 09:52:11.7 & 47:16:21 & 0.6259 & 0.0001 \\
  2 $^{(*)}$ & 09:52:12.8 & 47:16:39 & 0.6470 & 0.0004 \\
  3 $^{(*)}$ & 09:52:14.2 & 47:17:12 & 0.6492 & 0.0004 \\
  4 $^{(+)}$ & 09:52:12.9 & 47:17:29 & 0.6240 & 0.0001 \\
  5 $^{(+)}$ & 09:52:05.6 & 47:15:50 & 0.6356 & 0.0003 \\
  6 $^{(+)}$ & 09:52:18.3 & 47:14:26 & 0.6559 & 0.0005 \\
  7 $^{(+)}$ & 09:52:00.7 & 47:16:12 & 0.6502 & 0.0004 \\
  8 $^{(+)}$ & 09:51:49.2 & 47:14:18 & 0.6312 & 0.0002 \\
  9 $^{(+)}$ & 09:51:40.1 & 47:14:09 & 0.6307 & 0.0002 \\
 10 $^{(+)}$ & 09:52:50.6 & 47:13:37 & 0.6478 & 0.0001 \\
\hline
\multicolumn{5}{c}{Cluster P36} \\
  1 & 09:53:53.9 & 47:40:11 & 0.2459 & 0.0002 \\
  2 & 09:53:54.0 & 47:40:24 & 0.2502 & 0.0001 \\
  3 & 09:53:54.2 & 47:40:36 & 0.2484 & 0.0002 \\
\hline 
\multicolumn{5}{c}{Cluster P63} \\
  1 & 13:24:20.8 & 30:12:42 & 0.6870 & 0.0003 \\
  2 & 13:24:21.5 & 30:13:01 & 0.6866 & 0.0009 \\
\hline
\end{tabular}
\begin{list}{}{}
\item[$^{(*)}$] Two redshifts: from our TNG observations and from CFHT
archival data. The discrepancy between measurements is less than 50
km s$^{-1}$.
\item[$^{(+)}$] Redshift from CFHT archival data.
\end{list}
\end{table}

The spectroscopic survey of cluster V04 (P04) reveals a peak of 9
galaxies in the range $163000 < cz < 167000$ km s$^{-1}$.  Applying
the biweight estimator (Beers et al. \cite{Beers90}), we find that the
cluster mean redshift is $z=0.5505 \pm 0.0004$. The 9 member galaxies
are within $3'$, that is, within 1.9$\,h^{-1}$ Mpc from the center of
the cluster.  Based on these 9 galaxies, we compute a velocity
dispersion $\sigma_{\bf{v}}=311_{-97}^{+171}$ in the cluster rest frame
(Harrison \& Noonan \cite{Harrison79}).

This cluster has been identified by both us and PLG96: its
spectroscopic confirmation is reassuring. In particular it shows that
even relatively poor clusters can be detected with either the VGCF or
the matched-filter algorithm. Because we find no ETS in the CMD of
this cluster, this cluster also clarifies a limit of our color
confirmation technique: an ETS is clearly identifiable only in
relatively rich clusters.

In the case of cluster V06 we find a peak along the line-of-sight and
within $3'$ (1.3$\,h^{-1}$ Mpc) from the center. There are 9 galaxies
in the range $107000 < cz < 112000$ km s$^{-1}$. Using the biweight
estimator, we find that the cluster mean redshift is $z=0.366 \pm
0.001$. Its velocity dispersion is $\sigma_{\bf{v}}=797_{-238}^{+408}$
km s$^{-1}$ in the cluster rest frame.

The fact that we identify an ETS for V06 is consistent with the
cluster velocity dispersion, typical of a richer system than V04. The
(average) photometric redshift we obtain from the ETS is
$z_{\bf{phot}}=0.43$. There is a $\delta z \simeq 0.06$ difference between
the photometric and spectroscopic redshifts, well within our estimated
uncertainty of $\delta z \simeq 0.1$.  The spectroscopic confirmation
that V06 is a real physical system is important for the VGCF since V06
has not been identified by the matched-filter algorithm.

We are not able to identify a system along the line of sight of
V18. Redshifts are spread over a wide redshift range and show no
significant peak.  The result of the spectroscopic observations is
consistent with the fact that we do not find an ETS for V18.

We now turn to P33, P36 and P63 that are in PDCS regions not covered
by our run of the VGCF. The photometric and spectroscopic data we have
for these clusters provide a further test of our ability to identify
possible ETS and to estimate the photometric redshift of the cluster.

First of all we present in Fig. \ref{fig:CMR_PDCS333663} the CMD of
these clusters together with the ETS we identify following the same
procedure as in the previous section. The black circles are the
galaxies we use in the fit of the color-magnitude relation with fixed
slope (see Sect. \ref{sec:colorana}). Crosses are galaxies with
spectroscopic redshift: we use all these galaxies in the fit, but one:
the reddest galaxy in P33.

The photometric redshifts we derive for P33, P36, and P63 are
$z_{\bf{phot}} = 0.64 \pm 0.04$, $0.18 \pm 0.02$, and $0.56 \pm 0.06$,
respectively.

Cluster P33 presents a broad peak of 10 galaxies in redshift space.
The peak is at a mean redshift $z=0.640 \pm 0.004$, the same redshift
we estimate from the ETS.  The velocity dispersion we compute is
$\sigma_{\bf{v}}=2214_{-662}^{+1133}$ km s$^{-1}$ in the cluster rest
frame. This is an unrealistic velocity dispersion that could be caused
by a too sparse sampling of the velocity field of the cluster. We note
in fact that the redshift distribution is likely to be
double-peaked. If we split the redshift distribution into two
components at $z = 0.6290$ and $z = 0.6482$, we obtain two reasonable
velocity dispersions $\sigma_{\bf{v}}=569$ km s$^{-1}$ and $\sigma_{\bf{v}}=389$ km
s$^{-1}$.  We note that Holden et al. (\cite{Holden97}) find X--ray
emission at the position of P33 in the ROSAT All Sky Survey (Snowden
\& Schmitt \cite{Snowden90}). This X--ray emission could be explained 
by a merging scenario between the two clumps (see e.g. Barrena et al.
\cite{Barrena02} for a possible similar scenario in 1E0657-56 cluster of 
galaxies).

For clusters P36 and P63 we measure redshifts from long slit spectra:
three redshifts for P36 and two for P63. All the three galaxies of
cluster P36 have redshifts close to $z=0.248$ and both galaxies of P63
are at $z=0.686$. Table \ref{tab:our_redshifts} lists the redshifts of
these cluster members. Again, the spectroscopic redshifts are quite
close to the photometric estimates based on the ETS: ($z_{\bf{phot}} - z$)
= 0.07 and 0.13, respectively.

Redshifts we retrieve from the Sloan Digital Sky Survey team (SDSS;
see, e.g., York et al. \cite{York00}) provide further evidence that
the ETS is a reasonable indicator of the distance of our clusters. We
find redshifts of 4 galaxies, one for each of the following clusters
with ETS: V17, V23, V30, and V48.

Although one galaxy is not enough for a reliable redshift assignment,
our galaxies with redshift have a rather large probability of being
cluster members since they are among the brightest galaxies among 2D
members and lie at the center of the angular distribution of counts.
We list the redshifts in Table \ref{tab:redshiftSDSS}. For 3 out of 4
clusters with ETS we find $z_{\bf{phot}} - z < 0.02$.  Cluster V17 has
$z_{\bf{phot}} - z = 0.08$.

Finally, we consider the spectroscopic redshifts of Holden et al.
(\cite{Holden99}) for clusters P11, P12 and P23.  Holden et al.
\cite{Holden99} estimate for these 3 clusters $z=0.327 \pm 0.003$,
$z=0.263 \pm 0.002$, and $z=0.129 \pm 0.001$, respectively. We compare
these values with our photometric redshift estimates: $z$ = 0.36,
0.32, and 0.19, respectively.  The difference between photometric and
spectroscopic redshifts is $z_{\bf{phot}} - z < 0.06$.

In conclusion, we find that spectroscopic and photometric redshifts
are in good agreement. Table \ref{tab:photspecred} summarizes the
results obtained with both techniques. The average difference between
photometric and spectroscopic redshifts is $<z_{\bf{phot}}-z_{\bf{spec}}> = 0.05
\pm 0.04$, consistent with our estimate of external errors $\delta z
\simeq 0.1$ (see Sect. \ref{sec:colorana}).

\begin{table}
\caption{Redshifts obtained from SDSS}
\label{tab:redshiftSDSS}
\centering
\begin{tabular}{ccccc} 
\hline \hline
 Cluster & $\alpha_{\rm 2000}$ & $\delta_{\rm 2000}$ & $z$ & $\delta z$ \\
    & (hh:mm:ss) & ($^{\circ}$:$'$:$''$) &   &  \\
\hline
 Galaxy in V17 & 02:27:45 & 00:59:20 & 0.3965 & 0.0002 \\
 Galaxy in V23 & 02:28:33 & 00:57:18 & 0.3371 & 0.0002 \\
 Galaxy in V30 & 02:26:24 & 01:00:40 & 0.3979 & 0.0002 \\
 Galaxy in V48 & 02:25:23 & 01:08:50 & 0.2674 & 0.0002 \\
\hline
\end{tabular}
\end{table}

\section{Discussion of the VGCF and PDCS cluster detections}
\label{sec:discCMD}

The basic characteristics of the VGCF and matched-filter algorithm
(PLG96) are the numbers of clusters they detect. Table
\ref{tab:resumen} summarizes the situation. We identify a total of 48
clusters, including 3 possible substructures. 25 of the 48 clusters are
particularly reliable since we identify an ETS in their CMD. Given
that several clusters are rather poor, we point out that the lack of
an ETS is not to be taken as a proof that the cluster is in fact a
mere projection of unrelated galaxies.

Within the PDCS borders we identify 41 clusters, i.e. 13 more clusters
than PLG96 (all the 25 reliable clusters are among these 41
clusters). Because only a fraction of our 41 clusters coincide with
PDCS clusters, the difference between the numbers of VGCF and PDCS clusters
does not reflect the actual performances of the two algorithms. 
A more detailed discussion is needed in order to understand how 
complete the VGCF and PDCS catalogs are.

We start from Fig. \ref{fig:clust_tot}.  As already described in
Sect. {\ref{sec:clus_catalog}, in Fig. \ref{fig:clust_tot} our
clusters are the solid thick circles, PDCS clusters are the
long-dashed circles marked with a P followed by the PDCS
identification number. For an easier comparison, in what follows, we
consider only VGCF clusters that fall within the PDCS0 and PDCS2
areas.

There are 20 clusters that are identified by both the VGCF and the
matched-filter (PLG96). These ``common'' clusters have properties
that are representative of their parent samples in these fields.

We then identify 21 clusters that PLG96 miss. These clusters fall
into three categories: a) clusters that are identified within larger
clusters, b) clusters that are very close to other (larger) clusters,
and c) apparently ``normal'' isolated clusters.

Clusters V12, V14, and V22 are detected within clusters V13, V33, and
V21 respectively.  These clusters could have not been detected by the
matched-filter algorithm.  These (poor) clusters could be physical
substructures or projected background clusters.  We find no ETS in
their CMD. The same is true for V22 that lies inside PDCS2 but is
included within V21 identified outside PDCS2.

The category of small clusters identified close to larger clusters
includes 6 systems (V11, V15, V16, V17, V24, V41). In detail, we
consider as ``close'' clusters with centers at a distance $D_{12}$
such that $ max(R_1,R_2) < D_{12} < R_1 + R_2$, where $R_1$ and $R_2$
are the radii of the two clusters.

Clusters in this category mark a significant difference between VGCF
and matched-filter techniques.  For 4 out of 6 clusters we identify a
sequence of early-type galaxies that make these detection particularly
reliable.  The origin of this difference is likely to be the smoothing
performed by the matched filter algorithm.  The relatively large
smoothing length used for the detection of rich and/or low-redshift
clusters does not allow the detection of small neighboring structures.

Finally, we identify 12 well isolated clusters missing in the PDCS
catalog (V05, V06, V16, V18, V20, V31, V34, V37, V39,V44, V45, V47).
No obvious property of these clusters explains their absence in the
PDCS catalog, except (possibly) for cluster V30 that is only partially
contained within the PDCS field.

We find an ETS in the CMD of 6 out of the 12 well isolated clusters.  
Spectroscopic observations confirm cluster V06 (among the 6 with ETS) and add V04
(no ETS) to the list of confirmed clusters.  The relatively large
fraction of "color confirmations" and the two spectroscopic
confirmations, lead us to the conclusion that VGCF clusters
significantly contribute to the completeness of cluster samples
identified with a matched-filter algorithm.

PLG96 identify 8 clusters that the VGCF misses. Of these 8
clusters, only 2 are in PDCS0. All other clusters are in PDCS2. In
fact all but one of these clusters are located into the right
ascension range $2^{\bf{h}} 26^{\bf{m}} < \alpha < 2^{\bf{h}}
31^{\bf{m}}$, less than 1/3 of the total area of PDCS2 field. The
clusters with $2^{\bf{h}} 26^{\bf{m}} < \alpha < 2^{\bf{h}}
31^{\bf{m}}$ are P15, P17, P25, P26, P27. We identify clusters P26 and
P27 only in the V band because we have no I-band data at their
position.  These clusters do not enter our catalog since we require
the detection in two bands for a cluster.  We identify P15, P17, and
P25 only in the V band too, even if we have I-band data at their
position. We also note that P15 and P27 have the smallest radii among
the PDCS clusters in PDCS0 and PDCS2.

In conclusion, there are 46 clusters identified by us and/or PLG96
within the two PDCS fields (excluding the 3 possible
``substructures'' V12, V14, and V22). The total number of clusters is larger than the
number of VGCF clusters identified within the same region by 20\% and
larger than the number of PDCS clusters by more than 60\%.

\begin{table}
\caption{Cluster Counts}
\label{tab:resumen}
\centering
\begin{tabular}{lccl} 
\hline \hline
 & Confirmed & Not Confirmed & Total \\
\hline
\multicolumn{4}{c}{Within the total $BR$ + $VI$ area} \\
VGCF &     25    &     23        & 48 \\
\hline
\multicolumn{4}{c}{Within PDCS0 and PDCS2} \\
VGCF &     25    &     15        & 41 \\
PLG96       &     --    &     28        &  28          \\
\hline
\end{tabular}
\end{table}
 
As a final remark, we note that the comparison between our catalog and
the PDCS catalog can be extended to redshift estimates for 11 clusters
with redshift estimated by both PLG96 and us. We plot these
systems in Fig. \ref{fig:redshifts_xy}. In the same figure we draw a
diagonal line representing a one-to-one relation. For all but one
system (V03 = P01) the redshift estimates agree well within the
specified uncertainties, $\delta z = 0.1$ for both us and PLG96.

\begin{figure}
\centering
\includegraphics[width=8.8cm]{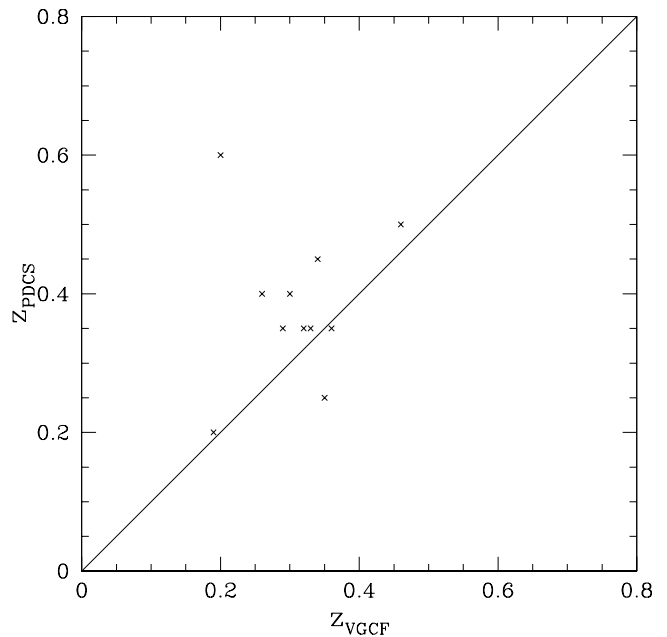}
\caption{Comparison between VGCF and PDCS redshift estimates.  The
diagonal line illustrates a one-to-one relation and it is not a fit.}
\label{fig:redshifts_xy}
\end{figure}

\begin{table}
\caption{Clusters with both photometric and spectroscopic redshifts.}
\label{tab:photspecred}
\centering
\begin{tabular}{cccl} 
\hline \hline
 Cluster    & PDCS & $\overline{z}_{\bf{phot}}$ & \multicolumn{1}{c}{$z_{\bf{spec}}$}   \\
\hline
  V06       &  --  & 0.43 & $0.3658 \pm 0.0009$           \\
  V13      & P12 & 0.32 & $0.2630 \pm 0.0020$ $^{(1)}$  \\
  V17      &  --  & 0.32 & $0.3965 \pm 0.0002$ $^{(2)}$  \\
  V19      & P23 & 0.19 & $0.1290 \pm 0.0010$ $^{(1)}$  \\
  V23      & P20 & 0.34 & $0.3371 \pm 0.0002$ $^{(2)}$  \\
  V30      &  --  & 0.38 & $0.3979 \pm 0.0002$ $^{(2)}$  \\
  V36      & P11 & 0.36 & $0.3270 \pm 0.0030$ $^{(1)}$  \\
  V48      & P24 & 0.26 & $0.2674 \pm 0.0002$ $^{(2)}$  \\
  --        & P33 & 0.64 & $0.6400 \pm 0.0040$           \\
  --        & P36 & 0.18 & $0.2480 \pm 0.0020$           \\
  --        & P63 & 0.56 & $0.6863 \pm 0.0003$           \\
\hline
\end{tabular}
\begin{list}{}{}
\item[$^{(1)}$] From Holden et al. (\cite{Holden99}).
\item[$^{(2)}$] Redshift from SDSS data for a single galaxy that falls
into the brightest part of the ETS of the cluster.
\end{list}
\end{table}

\section{Summary}

Our aim is to test the performances of the VGCF on deep $B$, $R$ wide
field images of two PDCS fields (PDCS0 at 0$^{\bf{h}}$ and PDCS2 at
2$^{\bf{h}}$ ).  In a previous paper (R01) we run the VGCF on PDCS
catalogs of galaxies in the $V$ and $I$ bands (PLG96). Here we
want to improve the assessment of the merits of the VGCF and the PDCS 
catalogs (obtained with a matched filter algorithm) in further two 
PDCS fields. In these fields we also  have deep $B$ and $R$ wide field 
images at our disposal. By comparing the PDCS and VGCF catalogs, we
evaluate the relative incompleteness of the two catalogs. This
information is relevant to the cosmological abundance of clusters and to
the evaluation of selection biases of optical cluster samples.

We run the VGCF in magnitude bins on our $BVRI$ catalogs and obtain a
final list of 48 clusters.  Of these clusters, 3 are identified within
other clusters and could be either physical substructures or projected
background clusters.

We use our $BVRI$ photometry to find evidences that our clusters are
real physical systems.  We analyze the CMD of each cluster, and search
for the presence of an ETS. We consider these clusters as reliable
VGCF identifications and include them in the list of ``confirmed
clusters''

Once we identify an ETS, we estimate the photometric redshift of
the cluster by comparing the observed ETS to a grid of ETS 
obtained by redshifting the ETS of Coma (for the $BR$ CMD) and A118 
(for the $VI$ CMD). 

We find an ETS in both the ($B-R$,$R$) and ($V-I$,$I$) CMD of 7
clusters. There are further 18 clusters that have a recognizable ETS
in one of the two CMD. Redshift estimates range from $z_{\bf{ETS}}$ = 0.12
(V31) to $z_{\bf{ETS}}$ = 0.75 (V5) with an uncertainty of the order of
$\delta z = 0.1$.

For three of our VGCF clusters (V04, V06, V18) we obtain additional
spectroscopic data with our own observations with the DOLoRes
spectrograph at the TNG or from public databases. We confirm V04 and
V06 as real physical systems. Consistently with our expectations, V18
that appears to be a projection of physically unrelated galaxies has
no detectable ETS.

We then compare our cluster catalog with the PDCS catalog (PLG96).
There are 41 VGCF clusters in the PDCS0 and PDCS2 areas where
PLG96 identify 28 clusters.

A total of 20 clusters are identified by both the VGCF and the
matched-filter algorithm of PLG96.

We identify 21 clusters that PLG96 miss. These clusters fall into
three categories: a) 3 clusters that are identified within larger
clusters, b) 6 clusters that are very close to other (larger)
clusters, and c) 12 apparently ``normal'' isolated clusters. In
particular case b) clusters indicate the main draw-back of the
smoothing required by the matched-filter algorithm.

PLG96 identify 8 clusters that the VGCF misses. 

In conclusion, the VGCF identifies a large fraction of the PDCS
clusters ($\sim$ 70\%). For part of the PDCS fields we
only have $V$ and $I$ data. If we relax our criterium for cluster
identification in these areas and require only a detection in the $V$ band, 
the VGCF identifies $\sim$ 90\% of the PDCS clusters.

Adding together all the independent cluster identifications, the total 
number of clusters within PDCS0 and PDCS2 is 46. This number is $\sim$ 
20\% larger than the number of clusters identified with the VGCF and 
$\sim$ 60\% larger than the number of PDCS clusters.

These results confirm a) that the VGCF is a competitive algorithm for
the identification of clusters, b) that a combined catalog of
matched-filter and VGCF clusters constitutes a significant progress
toward a more complete selection of clusters from bidimensional
optical data.

\begin{acknowledgements}
We wish to thank Marc Postman, who kindly provided us $V$ and $I$
catalogs of the 0$^{\bf{h}}$ and 2$^{\bf{h}}$ PDCS fields and Alex
Oscoz for his help during the observations at the INT and useful
comments and suggestions.

This publication is based on observations made on the island of La
Palma with the Isaac Newton Telescope (INT) and Jacobus Kaptein
Telescope (JKT) both operated by the Isaac Newton Group, with the
Italian Telescopio Nazionale Galileo (TNG) operated by the Centro
Galileo Galilei dell'INAF (Istituto Nazionale di Astrofisica), and the
Nordic Optical Telescope (NOT) in the Spanish Observatorio del Roque
de los Muchachos of the Instituto de Astrofisica de Canarias.

This publication also makes use of data from the Sloan Digital Sky
Survey (SDSS) public archive. Funding for the creation and
distribution of the SDSS Archive has been provided by the Alfred
P. Sloan Foundation, the Participating Institutions, the National
Aeronautics and Space Administration, the National Science Foundation,
the U.S. Department of Energy, the Japanese Monbukagakusho, and the
Max Planck Society. The SDSS is managed by the Astrophysical Research
Consortium (ARC) for the Participating Institutions. The Participating
Institutions are The University of Chicago, Fermilab, the Institute
for Advanced Study, the Japan Participation Group, The Johns Hopkins
University, the Korean Scientist Group, Los Alamos National
Laboratory, the Max-Planck-Institute for Astronomy (MPIA), the
Max-Planck-Institute for Astrophysics (MPA), New Mexico State
University, University of Pittsburgh, Princeton University, the United
States Naval Observatory, and the University of Washington.

\end{acknowledgements}


\end{document}